\documentclass[twocolumn,showpacs,preprintnumbers,amsmath,amssymb]{revtex4}

\usepackage{graphicx}
\usepackage{dcolumn}
\usepackage{bm}

\usepackage{amssymb}
\usepackage{amsmath}

\begin{document}


\title{Thermodynamic constraints on a varying cosmological-constant-like term \\ from the holographic equipartition law with a power-law corrected entropy}

\author{Nobuyoshi {\sc Komatsu}}  \altaffiliation{E-mail: komatsu@se.kanazawa-u.ac.jp} 

\affiliation{Department of Mechanical Systems Engineering, Kanazawa University, 
                          Kakuma-machi, Kanazawa, Ishikawa 920-1192, Japan }
\date{\today}

\begin{abstract}

A power-law corrected entropy based on a quantum entanglement is considered to be a viable black-hole entropy.
In this study, as an alternative to Bekenstein--Hawking entropy, a power-law corrected entropy is applied to Padmanabhan's holographic equipartition law to thermodynamically examine an extra driving term in the cosmological equations for a flat Friedmann--Robertson--Walker universe at late times.
Deviations from the Bekenstein--Hawking entropy generate an extra driving term (proportional to the $\alpha$-th power of the Hubble parameter, where $\alpha$ is a dimensionless constant for the power-law correction) in the acceleration equation, which can be derived from the holographic equipartition law. 
Interestingly, the value of the extra driving term in the present model is constrained by the second law of thermodynamics. 
From the thermodynamic constraint, the order of the driving term is found to be consistent with the order of the cosmological constant measured by observations.
In addition, the driving term tends to be constant-like when $\alpha$ is small, i.e., when the deviation from the Bekenstein--Hawking entropy is small.

\end{abstract}


\maketitle

\section{Introduction} 
\label{Introduction}

An accelerated expansion of the late universe implies a positive cosmological constant $\Lambda$ in standard $\Lambda$CDM (lambda cold dark matter) models \cite{PERL1998_Riess1998,Planck2015}. 
However, the measured cosmological constant is much smaller than the theoretical value from quantum field theory \cite{Weinberg1989}. 
Various cosmological models \cite{Prigogine_1988-1989,Lima-Others1996-2016,Freese-Mimoso_2015,Sola_2009-2015,Nojiri2006,Bamba} have been proposed to resolve this problem with the cosmological constant.
In particular, thermodynamic scenarios have been examined using the holographic principle, which assumes that the information of the bulk is stored on the horizon \cite{Hooft-Bousso}.   
For example, cosmological equations in an FRW (Friedmann--Robertson--Walker) universe \cite{Sheykhi1,Sadjadi1} have been studied from the viewpoint of entropic forces \cite{Jacob1995,Padma1,Verlinde1}, while an entropic cosmology, which assumes the usually neglected surface terms on the horizon, has been suggested \cite{Easson12,Koivisto1Basilakos1-Gohar,Koma4,Koma5-8,Koma9}.  
In these studies, the Bekenstein--Hawking entropy \cite{Bekenstein1Hawking1} is generally used, replacing the horizon of a black hole by the horizon of the universe.

Recently, another thermodynamic scenario called the `holographic equipartition law' has been proposed \cite{Padma2012A},
for which cosmological equations in a flat FRW universe were successfully derived using the Bekenstein--Hawking entropy from the expansion of cosmic space due to the difference between the degrees of freedom on the surface and in the bulk. 
The emergence of the cosmological equations has been examined from various viewpoints \cite{Padma2012,Cai2012-Tu2013,Tu2013-2015,Padma2014-2015,ZLWang2015,Tu2015,Yuan2013,Krishna2017}. 
However, an extra driving term, related to $\Lambda$ and a time varying $\Lambda (t)$, has not been directly derived from the Bekenstein--Hawking entropy \cite{Koma10}.

The Bekenstein--Hawking entropy is considered to be additive based on Boltzmann--Gibbs statistics.
However, self-gravitating systems exhibit peculiar features, such as nonextensive statistics \cite{Tsa1,Tsa0,Ren1}.
Therefore, the Tsallis--Cirto entropy \cite{Tsallis2012} and a modified R\'{e}nyi entropy \cite{Czinner1,Czinner2} have been proposed for black-hole entropy.  
In fact, the present author has applied the modified R\'{e}nyi entropy to the holographic equipartition law \cite{Koma10}.
Consequently, a deviation from the Bekenstein--Hawking entropy is found to play an important role. 
In particular, a constant-like driving term can be derived when a specific condition is mathematically satisfied \cite{Koma10};
however, the mathematical condition cannot be explained from a physical viewpoint.
In addition, the physical origin of the modified R\'{e}nyi entropy is unclear \cite{Koma10}.
Therefore, it is worthwhile using an alternative type of entropy to develop a deeper understanding of the extra driving term in cosmological equations from the holographic equipartition law.

For example, quantum corrections, such as logarithmic corrections and power-law corrections, have been proposed for black-hole entropy. 
The logarithmic correction arises from loop quantum gravity \cite{LQG2004_1,LQG2004_2,LQG2004_3}, while the power-law correction appears in treatments of the entanglement of quantum fields between the inside and outside of the horizon \cite{Das2008,Radicella2010}.
The present author has found that the power-law corrected entropy \cite{Das2008} is suitable for the holographic equipartition law, due to both its power-law formula and small corrections at late times.
In fact, the power-law corrected entropy has been used to study the generalized second law in universes \cite{Radicella2010,Sheykhi2,Karami2011}, holographic dark energy models \cite{Sheykhi2011}, $f(T)$-gravity models \cite{Karami2013,Saha2016}, and other applications.

In this context, we apply the power-law corrected entropy to the holographic equipartition law to thermodynamically examine an extra driving term in cosmological equations in a flat FRW universe.
The value of the driving term is expected to be restricted by the second law of thermodynamics.
Through the present study, the order of the extra driving term can be discussed from a thermodynamics viewpoint.

The remainder of the present article is organized as follows.
In Sec.\ \ref{Entropy on the horizon}, entropies on the Hubble horizon of a flat FRW universe are discussed.
The Bekenstein--Hawking entropy is reviewed in Sec.\ \ref{Bekenstein-Hawking entropy}, while a power-law corrected entropy is introduced in Sec.\ \ref{Power-law}.
Padmanabhan's holographic equipartition law is briefly reviewed in Sec.\ \ref{Holographic equipartition}.
In Sec.\ \ref{Power-law entropy and holographic equipartition law}, the power-law corrected entropy is applied to the holographic equipartition law, to derive an acceleration equation that includes an extra driving term.
In Sec.\ \ref{GSL}, the generalized second law of thermodynamics for the present model is examined and the order of the driving term is discussed from a thermodynamics viewpoint.
Finally, in Sec.\ \ref{Conclusions}, the conclusions of the study are presented.

It should be noted that an assumption of equipartition of energy used for the holographic equipartition law has not yet been established in a cosmological spacetime.
In addition, a power-law corrected entropy looks strange, due to its power-law formula.
However, the holographic equipartition law with the power-law corrected entropy is expected to play an important role in examining an extra driving term in cosmological equations thermodynamically.
Therefore, in this paper, as a viable scenario, the power-law corrected entropy is applied to the holographic equipartition law.

\section{Entropy on the Hubble horizon} 
\label{Entropy on the horizon}

The Bekenstein--Hawking entropy \cite{Bekenstein1Hawking1} is generally used as an associate entropy on the horizon of the universe.
Accordingly, in Sec.\ \ref{Bekenstein-Hawking entropy}, the Bekenstein--Hawking entropy is briefly reviewed. 
In Sec.\ \ref{Power-law}, a power-law corrected entropy \cite{Das2008} is introduced.
In the present study, the power-law corrected entropy is used for entropy on the Hubble horizon in a flat FRW universe, because its power-law formula is suitable for the holographic equipartition law.
Logarithmic corrections based on loop quantum gravity \cite{LQG2004_1,LQG2004_2,LQG2004_3} are not discussed in this paper.

\subsection{ Bekenstein--Hawking entropy} 
\label{Bekenstein-Hawking entropy} 

The Bekenstein--Hawking entropy $S_{\rm{BH}}$ is written as
\begin{equation}
 S_{\rm{BH}}  = \frac{ k_{B} c^3 }{  \hbar G }  \frac{A_{H}}{4}   ,
\label{eq:SBH}
\end{equation}
where $k_{B}$, $c$, $G$, and $\hbar$ are the Boltzmann constant, the speed of light, the gravitational constant, and the reduced Planck constant, respectively \cite{Bekenstein1Hawking1}. 
The reduced Planck constant is defined by $\hbar \equiv h/(2 \pi)$, where $h$ is the Planck constant. 
$A_{H}$ is the surface area of the sphere with the Hubble horizon (radius) $r_{H}$, given by
\begin{equation}
     r_{H} = \frac{c}{H}   , 
\label{eq:rH}
\end{equation}
where the Hubble parameter $H$ is defined by \cite{Koma4,Koma5-8,Koma9} 
\begin{equation}
   H \equiv   \frac{ da/dt }{a(t)} =   \frac{ \dot{a}(t) } {a(t)}  , 
\label{eq:Hubble}
\end{equation}
and  $a(t)$ is the scale factor at time $t$.
Substituting $A_{H}=4 \pi r_{H}^2 $ into Eq.\ (\ref{eq:SBH}) and using Eq.\ (\ref{eq:rH}), we obtain \cite{Koma4,Koma5-8,Koma9,Koma10} 
\begin{equation}
S_{\rm{BH}}  = \frac{ k_{B} c^3 }{  \hbar G }   \frac{A_{H}}{4}       
                  =  \left ( \frac{ \pi k_{B} c^5 }{ \hbar G } \right )  \frac{1}{H^2}  
                  =    \frac{K}{H^2}    , 
\label{eq:SBH2}      
\end{equation}
where $K$ is a positive constant given by
\begin{equation}
  K =  \frac{  \pi  k_{B}  c^5 }{ \hbar G } = \frac{  \pi  k_{B}  c^2 }{ L_{p}^{2} } , 
\label{eq:K-def}
\end{equation}
and $L_{p}$ is the Planck length, written as
\begin{equation}
  L_{p} = \sqrt{ \frac{\hbar G} { c^{3} } } .
\label{eq:Lp}
\end{equation}

From Eq.\ (\ref{eq:SBH2}), the rate of change of entropy is given by  
\begin{equation}
\dot{S}_{\rm{BH}}  =  \frac{d}{dt} \left ( \frac{K}{H^{2}} \right )  =  \frac{-2K \dot{H} }{H^{3}}   .  
\label{eq:dSBH}      
\end{equation}
Numerous observations imply $H >0$ and $\dot{H} < 0$ \cite{Krishna2017}.
(For observed data, see, e.g., Ref.\ \cite{Farooq2017}.) 
Therefore, the second law of thermodynamics for the Bekenstein--Hawking entropy should satisfy
\begin{equation}
\dot{S}_{\rm{BH}}   =  \frac{-2K \dot{H} }{H^{3}}  >  0  .  
\label{eq:dSBH_2}      
\end{equation}

\subsection{Power-law corrected entropy} 
\label{Power-law} 

Das \textit{et al.} have suggested a power-law corrected entropy, based on the entanglement of quantum fields between the inside and outside of the horizon \cite{Das2008}.
The formula of the power-law corrected entropy is summarized in the work of Radicella and Pav\'{o}n \cite{Radicella2010}. 
In addition, Sheykhi and Hendi have pointed out that power-law corrections are expected to be small in the late universe, whereas the corrections are large in the early universe \cite{Sheykhi2}.

According to Ref.\ \cite{Radicella2010}, the power-law corrected entropy can be written as
\begin{equation}
S_{pl}  = S_{\rm{BH}}  \left ( 1- K_{\alpha} A_{H}^{1- \frac{\alpha}{2}}  \right )   , 
\label{eq:Spl}      
\end{equation}
where $\alpha$ is a dimensionless parameter and $K_{\alpha}$ is given by
\begin{equation}
  K_{\alpha}   = \frac{  \alpha ( 4\pi )^{ \frac{\alpha}{2}-1} }{(4-\alpha)r_{c}^{2-\alpha} }   , 
 \label{eq:Ka}      
\end{equation}
and $r_{c}$ is the crossover scale. When $\alpha =0$, $S_{pl}$ becomes $S_{\rm{BH}}$. 
Note that $K_{\alpha}$ is different from $K$ in Eq.\ (\ref{eq:K-def}). 

Substituting Eq.\ (\ref{eq:Ka}) and $A_{H}=4 \pi r_{H}^2 $ into Eq.\ (\ref{eq:Spl}) and using $r_{H} = c/H$ and $r_{H0} = c/H_{0}$, we obtain 
\begin{align}
S_{pl}  &= S_{\rm{BH}}  \left [ 1- \frac{  \alpha ( 4\pi )^{ \frac{\alpha}{2}-1} }{(4-\alpha)r_{c}^{2-\alpha} }  (4 \pi r_{H}^2)^{1- \frac{\alpha}{2}}  \right ]  \notag \\   
           &= S_{\rm{BH}}  \left [ 1- \frac{\alpha}{4-\alpha}  \left ( \frac{r_{H}}{r_{c}}  \right )^{2- \alpha}  \right ]    \notag \\
           &= S_{\rm{BH}}  \left [ 1- \frac{\alpha}{4-\alpha}  \left ( \frac{r_{H0}}{r_{c}} \frac{H_{0}}{H}  \right )^{2- \alpha}  \right ]   , 
\label{eq:Spl2}      
\end{align}
where $H_{0}$ and $r_{H0}$ are the Hubble parameter and the Hubble radius at the present time, respectively.
Therefore, from Eq.\ (\ref{eq:Spl2}), the entropy $S_{H}$ on the horizon can be rewritten as
\begin{equation}
S_{H} = S_{pl}  = S_{\rm{BH}}  \left [ 1-  \Psi_{\alpha} \left ( \frac{H_{0}}{H} \right )^{2- \alpha}  \right ]    , 
\label{eq:SH2}      
\end{equation}
where $\Psi_{\alpha}$ is a dimensionless parameter given by 
\begin{equation}
  \Psi_{\alpha}    = \frac{\alpha}{4-\alpha} \left ( \frac{r_{H0}}{r_{c}} \right )^{2-\alpha}   , 
\label{eq:psi_a}      
\end{equation}
and $\alpha$ and $\Psi_{\alpha}$ are considered to be constant. 
In this study, $\Psi_{\alpha}$ is assumed to be positive for an accelerating universe, as examined in Sec.\ \ref{Power-law entropy and holographic equipartition law}.
Thus, $0 < \alpha <4$ is obtained from Eq.\ (\ref{eq:psi_a}) and $\Psi_{\alpha}>0$.
(The crossover scale $r_c$ can likely be identified with $r_{H0}$ \cite{Radicella2010,Dvali2003}.
In this case, $\Psi_{\alpha}$ reduces to $\frac{\alpha}{4-\alpha}$.)

To discuss the second law of thermodynamics, we examine the rate of change of $S_{H}$.
Differentiating Eq.\ (\ref{eq:SH2}) with respect to $t$ and using Eqs.\ (\ref{eq:SBH2}) and (\ref{eq:dSBH}), we have  
\begin{align}
\dot{S}_{H}  
                 &=  \dot{S}_{\rm{BH}}  \left [  1-   \frac{\Psi_{\alpha} H_{0}^{2- \alpha} }{H^{2- \alpha} } \right ]  + {S}_{\rm{BH}}  \left [ \frac{ (2-\alpha) \Psi_{\alpha} H_{0}^{2-\alpha} \dot{H} }{H^{3-\alpha}}  \right ] \notag \\
                 &=  \frac{-2K \dot{H} }{H^{3}} \left [  1-   \frac{\Psi_{\alpha} H_{0}^{2- \alpha} }{H^{2- \alpha} }  \right ]  +  \frac{K}{H^{2}}  \frac{ 2(1-\frac{\alpha}{2}) \Psi_{\alpha} H_{0}^{2-\alpha} \dot{H} }{H^{3-\alpha}}   \notag \\
                 &=  \frac{-2K \dot{H} }{H^{3}} \left [ 1-  \frac{ (2- \frac{\alpha}{2}) \Psi_{\alpha} H_{0}^{2-\alpha}}{H^{2-\alpha}}  \right ]    \notag \\
                 &=  \dot{S}_{\rm{BH}}  \left [ 1-  \left ( \frac{4- \alpha}{2}  \right ) \Psi_{\alpha}  \left ( \frac{H_{0}}{H} \right )^{2-\alpha}  \right ]       ,
\label{eq:dSH}      
\end{align}
where $\dot{S}_{\rm{BH}} >  0$ from Eq. (\ref{eq:dSBH_2}).
Accordingly, to satisfy $\dot{S}_{H} >  0 $, we require 
\begin{equation}
 1-  \left ( \frac{4-\alpha}{2} \right )  \Psi_{\alpha}  \left ( \frac{H_{0}}{H} \right )^{2-\alpha}    >  0    .
\label{eq:dSH_ineq}      
\end{equation}
Substituting Eq.\ (\ref{eq:psi_a}) into Eq.\ (\ref{eq:dSH_ineq}) and using $r_{H} = c/H$ and $r_{H0} = c/H_{0}$, we get an equivalent inequality:
\begin{equation}
 1-  \frac{\alpha}{2}   \left ( \frac{r_H}{r_c} \right )^{2-\alpha}     >  0    .
\label{eq:dSH_ineq00}      
\end{equation}
Using these constraints, we can discuss the order of an extra driving term in cosmological equations, as examined later. 
The inequality given by Eq.\ (\ref{eq:dSH_ineq00}) is consistent with that in Ref.\ \cite{Sheykhi2}.
If $r_c = r_{H}$, Eq.\ (\ref{eq:dSH_ineq00}) reduces to $\alpha < 2$, which is consistent with the result in Refs.\ \cite{Karami2011,Sheykhi2}.
That is, the constraint on a positive $\alpha$ can be written as $0 < \alpha <2$.
This constraint is stricter than the previously mentioned one, $0 < \alpha < 4$, which is related to an accelerating universe.
Therefore, the strict constraint, $0 < \alpha < 2$, is used here.
(The generalized second law of thermodynamics is examined in Refs.\ \cite{Karami2011,Sheykhi2}.
Those works are discussed in Sec.\ \ref{GSL}.)

The power-law corrected entropy looks strange due to its power-law formula although it is considered to be a viable black-hole entropy. 
However, as examined in Sec.\ \ref{Power-law entropy and holographic equipartition law}, both its power-law formula and small corrections at late times are suitable for the holographic equipartition law.
(The small correction could be interpreted as a weak quantum entanglement in the late universe.)
Accordingly, in this paper, the power-law corrected entropy is applied to the holographic equipartition law.

In the present study, we consider an entropy on the Hubble horizon of a flat FRW universe.
In a flat universe ($k = 0$), the Hubble horizon $r_{H} =c/H$ is equivalent to an apparent horizon $r_{A}$, because $r_{A}$ is given by $r_{A} = c/\sqrt{H^2 + (k /a^2) }$, where $k$ is a curvature constant.
A non-flat universe for the holographic equipartition law is examined in Ref.\ \cite{Cai2012-Tu2013} and an apparent horizon for power-law corrections is studied in Refs.\ \cite{Sheykhi2,Radicella2010,Karami2011}.

\section{Holographic equipartition law} 
\label{Holographic equipartition}

In this section, Padmanabhan's holographic equipartition law is introduced \cite{Padma2012A}.
A brief review of the law is also given in my previous study \cite{Koma10}, based on Padmanabhan's work \cite{Padma2012A} and other related works \cite{Padma2012,Cai2012-Tu2013,Tu2013-2015,Padma2014-2015,ZLWang2015,Tu2015}.

In an infinitesimal interval $dt$ of cosmic time, the increase $dV$ of the cosmic volume can be expressed as 
\begin{equation}
     \frac{dV}{dt}  =  L_{p}^{2} (N_{\rm{sur}} - \epsilon N_{\rm{bulk}} ) \times c      , 
\label{dVdt_N-N}
\end{equation}
where $N_{\rm{sur}}$ is the number of degrees of freedom on a spherical surface of Hubble radius $r_{H}$, while $N_{\rm{bulk}}$ is the number of degrees of freedom in the bulk \cite{Padma2012A}. 
$L_{p}$ is the Planck length given by Eq.\ (\ref{eq:Lp}) and $\epsilon$ is a parameter discussed below.
In the present study, $c$ is not set to be $1$ and, therefore, the right-hand side of Eq.\ (\ref{dVdt_N-N}) includes $c$ \cite{Koma10}. 
Using $r_{H}= c/H$, the Hubble volume $V$ can be written as
\begin{equation}
V = \frac{4 \pi}{3} r_{H}^{3} =  \frac{4 \pi}{3} \left ( \frac{c}{H} \right )^{3}   .
\label{eq:V}
\end{equation}
From Eq.\ (\ref{eq:V}), the rate of change of volume is given by
\begin{equation}
     \frac{dV}{dt}  =      \frac{d}{dt} \left ( \frac{4 \pi}{3} \left ( \frac{c}{H} \right )^{3}  \right )   =  -4 \pi c^{3}   \left (  \frac{ \dot{H} }{H^{4} } \right )  .
\label{dVdt_right}
\end{equation}
In this calculation, $r$ has been set to be $r_{H}$ before the time derivative is calculated \cite{Padma2012A}. 

The number of degrees of freedom in the bulk is assumed to obey the equipartition law of energy \cite{Padma2012A}: 
\begin{equation}
  N_{\rm{bulk}} = \frac{|E|}{ \frac{1}{2} k_{B} T}     , 
\label{N_bulk}
\end{equation}
where the Komar energy $|E|$ contained inside the Hubble volume $V$ is given by 
\begin{equation}
|E| =  |( \rho c^2 + 3p)| V  = - \epsilon ( \rho c^2 + 3p) V  ,
\label{Komar}
\end{equation}
and $\rho$ and $p$ are the mass density of cosmological fluids and the pressure of cosmological fluids, respectively \cite{Koma10}.
$\epsilon$ is a parameter defined as  \cite{Padma2012A,Padma2012}
 \begin{equation}
        \epsilon \equiv     
 \begin{cases}
              +1  & (\rho c^2 + 3p <0  \textrm{: an accelerating universe}),  \\ 
              -1  & (\rho c^2 + 3p >0   \textrm{: a decelerating universe}).    \\
\end{cases}
\label{epsilon}
\end{equation}
The temperature $T$ on the horizon is written as
\begin{equation}
 T = \frac{ \hbar H}{   2 \pi  k_{B}  }   .
\label{eq:T0}
\end{equation}
The number of degrees of freedom on the spherical surface is given by  
\begin{equation}
  N_{\rm{sur}} = \frac{4 S_{H} }{k_{B}}       , 
\label{N_sur}
\end{equation}
where $S_{H}$ is the entropy on the Hubble horizon \cite{Koma10}. 
Various types of entropy can be used for $S_{H}$.
In the next section, $S_{H}$ is set to be a power-law corrected entropy given by Eq.\ (\ref{eq:SH2}).

We now derive an acceleration equation from the holographic equipartition law.
In this paper, $\rho c^2 + 3p <0$ is selected \cite{Padma2012A} and, therefore, $\epsilon = +1$ from Eq.\ (\ref{epsilon}).
(The following result is not affected by this selection, because the same result can be obtained even if $\rho c^2 + 3p >0$ is selected \cite{Padma2012}.)
We first calculate $N_{\rm{bulk}}$ in the right-hand side of Eq.\ (\ref{dVdt_N-N}).
Substituting Eqs.\ (\ref{Komar}) and (\ref{eq:T0}) into Eq.\ (\ref{N_bulk}) and using Eq.\ (\ref{eq:V}), we obtain $N_{\rm{bulk}}$ given by \cite{Koma10}
\begin{align}
  N_{\rm{bulk}}  &= \frac{|E|}{ \frac{1}{2} k_{B} T}  
                       =   -  \frac{ (4 \pi)^{2} c^{5}  }{ 3 \hbar } \left (  \rho  + \frac{3p}{c^{2}}  \right )  \frac{1}{  H^{4}   }      .
\label{N_bulk_cal}
\end{align}
In addition, substituting $\epsilon = +1$ and Eqs.\ (\ref{eq:Lp}), (\ref{dVdt_right}), (\ref{N_sur}), and (\ref{N_bulk_cal}) into Eq.\ (\ref{dVdt_N-N}) and solving the resultant equation with regard to $\dot{H}$ \cite{Koma10}, we have 
\begin{align}
  \dot{H}                                                     
               &=   -  \frac{ 4 \pi G }{ 3} \left (  \rho  + \frac{3p}{c^{2}}  \right )  - \frac{ S_{H} H^{4} }{K}                        , 
\label{dVdt_N-N_cal2}
\end{align}
where $K$ is given by Eq.\ (\ref{eq:K-def}).
Substituting Eq.\ (\ref{dVdt_N-N_cal2}) into $ \ddot{a}/ a   =  \dot{H} + H^{2}$ and using $S_{\rm{BH}} = K/H^{2}$ given by Eq.\ (\ref{eq:SBH2}), the acceleration equation is written as
\begin{align}
  \frac{ \ddot{a} }{ a }       &=  \dot{H} + H^{2}             
                                      =   -  \frac{ 4 \pi G }{ 3} \left (  \rho  + \frac{3p}{c^{2}}  \right )  - \frac{ S_{H} H^{4} }{K}      + H^{2}  \notag \\
                                      &=   -  \frac{ 4 \pi G }{ 3} \left (  \rho  + \frac{3p}{c^{2}}  \right ) + H^{2}  \left (  1 - \frac{ S_{H} }{ S_{\rm{BH}} }  \right )     .
\label{N-N_FRW02_SH}
\end{align}
When $S_{H} = S_{\rm{BH}}$, the second term $H^{2}(1-S_{H}/S_{\rm{BH}})$ on the right-hand side is zero \cite{Padma2012A}.
However, when $S_{H} \neq S_{\rm{BH}}$, the second term is non-zero, i.e., an extra driving term appears.
The driving term depends on the deviation of $S_{H}$ from $S_{\rm{BH}}$.

As mentioned previously, an assumption of equipartition of energy is used for the holographic equipartition law, according to Ref.\ \cite{Padma2012A}.
However, the assumption has not yet been established in a cosmological spacetime.
This task is left for future research.
In the present study, the assumption of equipartition is considered to be a viable scenario even in the cosmological spacetime.

\section{Holographic equipartition law with a power-law corrected entropy} 
\label{Power-law entropy and holographic equipartition law}

In this section, a power-law corrected entropy given by Eq.\ (\ref{eq:SH2}) is applied to the holographic equipartition law.
The power-law corrected entropy is written as
\begin{equation}
S_{H} = S_{pl}  = S_{\rm{BH}}  \left [ 1-  \Psi_{\alpha} \left ( \frac{H_{0}}{H} \right )^{2- \alpha}  \right ]  ,
\label{eq:SH2_2}      
\end{equation}
where $\alpha$ and $\Psi_{\alpha}$ are dimensionless positive constants.
Substituting Eq.\ (\ref{eq:SH2_2}) into Eq.\ (\ref{N-N_FRW02_SH}), we have the acceleration equation:
\begin{equation}
  \frac{ \ddot{a} }{ a }           =   -  \frac{ 4 \pi G }{ 3} \left (  \rho  + \frac{3p}{c^{2}}  \right ) +  \Psi_{\alpha} H_{0}^{2- \alpha}  H^{\alpha}      .
\label{FRW02_SH0}
\end{equation}
The second term on the right-hand side, $\Psi_{\alpha} H_{0}^{2- \alpha}  H^{\alpha} $, is an extra driving term proportional to $H^{\alpha}$.
A positive second term is required for an accelerating universe.
Accordingly, $\Psi_{\alpha} >0$ is required because an expanding universe requires $H >0$.
Consequently, $0 < \alpha <4$ is obtained from Eq.\ (\ref{eq:psi_a}) and $\Psi_{\alpha} >0$.
(The above acceleration equation is different from that examined in previous studies  \cite{Tu2015,Koma10,Sheykhi2,Radicella2010,Sheykhi2011,Karami2011,Karami2013,Saha2016}.
A similar driving term was discussed in Ref.\ \cite{Yuan2013}.)

In this paper, the present model is considered to be a particular case of $\Lambda (t)$CDM models in time-varying $\Lambda(t)$ cosmologies \cite{Koma10}.
Accordingly, the Friedmann, acceleration, and continuity equations can be written as
\begin{equation}
    H^2     =  \frac{ 8\pi G }{ 3 } \rho  +  f_{\alpha} (H)  ,  
\label{FRW01_SH}
\end{equation}
\begin{equation}
  \frac{ \ddot{a} }{ a }           =   -  \frac{ 4 \pi G }{ 3} \left (  \rho  + \frac{3p}{c^{2}}  \right ) +   f_{\alpha} (H)  , 
\label{FRW02_SH}
\end{equation}
\begin{equation}
       \dot{\rho} + 3  \frac{\dot{a}}{a} \left (  \rho + \frac{p}{c^2}  \right )    =    - \frac{3 \dot{f}_{\alpha} (H) }{8 \pi G}       ,
\label{drho_SH}
\end{equation}
where $f_{\alpha} (H)$ is the extra driving term given by   
\begin{equation}
        f_{\alpha} (H) =  \Psi_{\alpha} H_{0}^{2- \alpha}  H^{\alpha}       .
\label{eq:fa}
\end{equation}
When $f_{\alpha} (H)$ is constant, the continuity equation is written as $ \dot{\rho} +  3 (\dot{a}/a) ( \rho   +  p/c^2 ) = 0 $, as for $\Lambda$CDM models.
In contrast, when $f_{\alpha} (H)$ varies with time, the right-hand side of the continuity equation  is nonzero.
In the holographic principle, the nonzero right-hand side can be interpreted as a kind of transfer of energy between the bulk (the universe) and the boundary (the horizon of the universe) \cite{Koma10}.
The energy transfer is expected to be small, as discussed later.

\begin{figure} [t]  
\begin{minipage}{0.495\textwidth}
\begin{center}
\scalebox{0.32}{\includegraphics{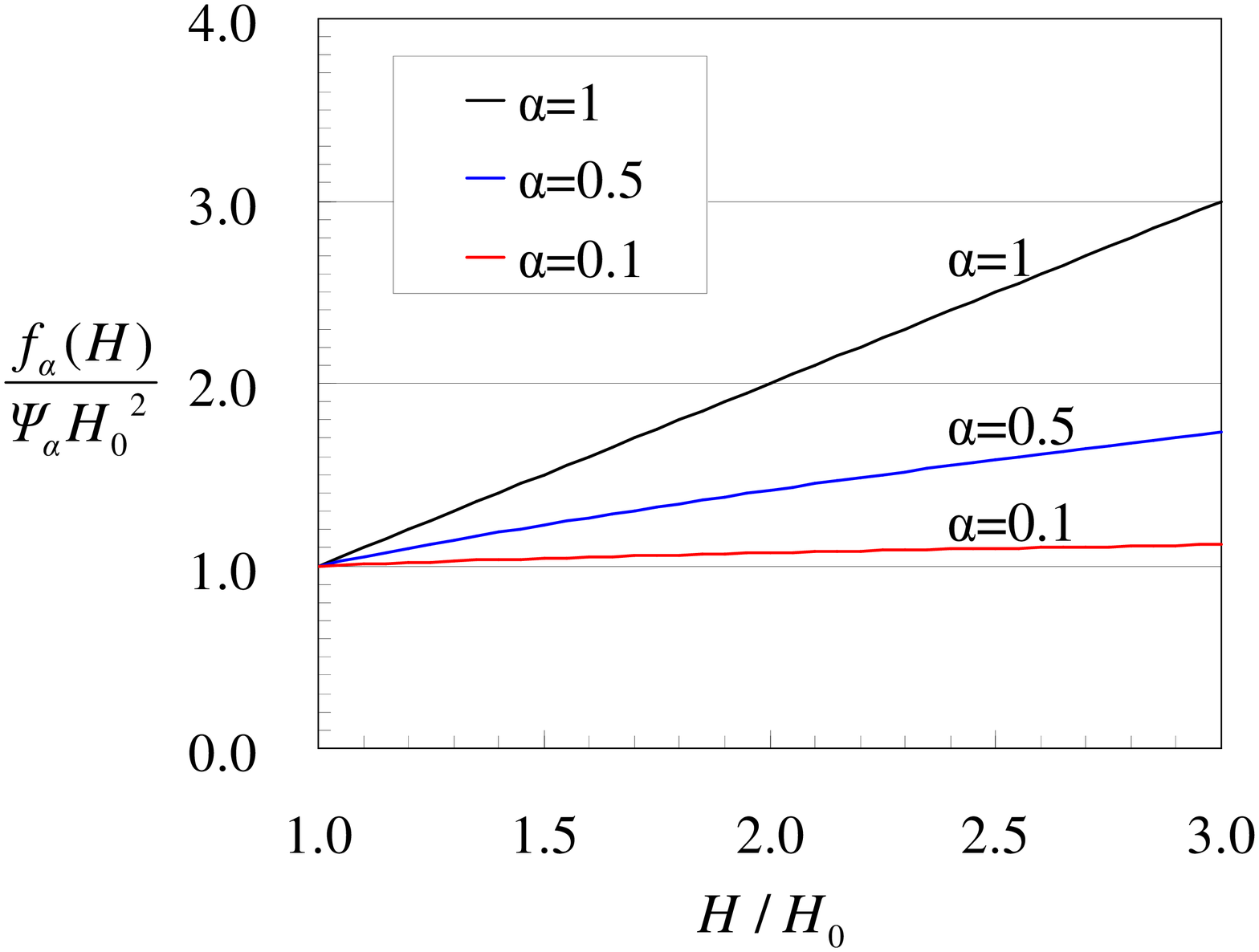}}
\end{center}
\end{minipage}
\caption{ (Color online). Evolutions of the normalized extra-driving term $f_{\alpha}(H)/(\Psi_{\alpha} H_{0}^{2})$ in the late universe.
 The normalized term is given by $(H/H_{0})^{\alpha}$, from Eq.\ (\ref{eq:fa2}). }
\label{Fig-H-H0-A}
\end{figure}

We now observe evolutions of $f_{\alpha}(H)$ in the late universe.
To this end, Eq.\ (\ref{eq:fa}) is rewritten as 
\begin{equation}
        f_{\alpha} (H) =  \Psi_{\alpha} H_{0}^{2- \alpha}  H^{\alpha}  =   \Psi_{\alpha} H_{0}^{2} \left (  \frac{H}{H_{0}} \right )^{\alpha}  , 
\label{eq:fa2}
\end{equation}
where $\Psi_{\alpha} H_{0}^{2}$ is constant.
The normalized extra-driving term $f_{\alpha}(H)/(\Psi_{\alpha} H_{0}^{2})$ given by $(H/H_{0})^{\alpha}$ is observed here.
Typical results for $\alpha =0.1$, $0.5$, and $1$ are plotted in Fig.\ \ref{Fig-H-H0-A}.
In this figure, $H/H_{0}$ varies from $1$ to $3$, which approximately corresponds to the range of redshift $z$ from $0$ to $2.4$ \cite{Farooq2017}. 
As shown in Fig.\ \ref{Fig-H-H0-A}, the normalized extra-driving term is not greatly influenced by $H/H_{0}$ when $\alpha$ is small, e.g., $\alpha =0.1$.
That is, the extra driving term tends to be constant-like for small values of $\alpha$.
In the next section, the order of the driving term is examined from a thermodynamics viewpoint.

As discussed in Ref.\ \cite{Sheykhi2}, power-law corrections based on the entanglement of quantum fields are expected to be small in the late universe, whereas they are large in the early universe.
Therefore, the small value of $\alpha$ could be interpreted as a weak entanglement in the late universe.
Of course, the present result depends on the choice of entropy.
Note that a similar constant-like term can be obtained when the deviation of a modified R\'{e}nyi entropy from the Bekenstein--Hawking entropy is small \cite{Koma10}.

The cosmological equations in the present model are considered to be equivalent to those in $\Lambda (t)$CDM models.
Various driving terms have been examined in the $\Lambda (t)$CDM model \cite{Freese-Mimoso_2015,Sola_2009-2015}.
In particular, a combination of the constant and $H^{2}$ terms, i.e., $\Lambda (t) = C_{0} H_{0}^{2} + C_{1} H^{2}$, was found to be favored \cite{Sola_2015L14},
where $C_{0}$ and $C_{1}$ are dimensionless constants.
For example, Sol\`{a} \textit{et al.} \cite{Sola_2015L14} have found that extra driving terms slightly deviate from a constant value because $C_{1} H^{2}$ terms exist and are small.
That is, the extra driving term is constant-like and the deviation from a constant is small.
Similarly, in the present model, the extra driving term, $f_{\alpha} (H)$, should be constant-like.
Consequently, the energy transfer across the horizon, related to the right-hand side of Eq.\ (\ref{drho_SH}), is expected to be small.

The background evolution for the present model can be calculated from Eqs.\ (\ref{FRW01_SH}), (\ref{FRW02_SH}), and (\ref{eq:fa}), using a method in Ref.\ \cite{Koma10}.
The solution is written as
\begin{equation}
       \frac{a}{a_{0}} =  \left [ \frac{  1- \Psi_{\alpha}  }{  (H/H_{0})^{2-\alpha} - \Psi_{\alpha}  } \right ]^{\frac{2}{3(2-\alpha)}}   , 
\label{eq:Sol_a_SH}
\end{equation}
where $\Psi_{\alpha}$ is given by Eq.\ (\ref{eq:psi_a}) and $a_{0}$ is the scale factor at the present time.
In the $\Lambda$CDM model, the background evolution is given by 
\begin{equation}
  \frac{a}{a_{0}} =  \left [ \frac{  1- \Omega_{\Lambda}  }{  (H/H_{0})^{2} - \Omega_{\Lambda}  } \right ]^{\frac{1}{3}}, 
\label{eq:Sol_a_LCDM}
\end{equation}
where $\Omega_{\Lambda}$ is the density parameter for $\Lambda$.
From Eqs.\ (\ref{eq:Sol_a_SH}) and (\ref{eq:Sol_a_LCDM}), it is found that when $\alpha$ is small, $\Psi_{\alpha}$ can behave as if it is $\Omega_{\Lambda}$.

The present model is considered to be a particular case of $\Lambda (t)$CDM models.
However, it may be possible to discuss this model from a different viewpoint.
For example, when $\alpha$ is small, a power-law corrected entropy given by Eq.\ (\ref{eq:Spl}) is approximately equivalent to $S_{\textrm{BH}}$.
In this case, it can be considered that a small deviation from $S_{\textrm{BH}}$ should be included in constants in Eq.\ (\ref{eq:SBH}), e.g., $G$ could be interpreted as a varying gravitational constant. 
That is, when $\alpha$ is small, the present model can behave as if it is a scalar field cosmology with a slowly varying action. 
For scalar field theories, see, e.g., Ref.\ \cite{Bamba} and references therein.
From this viewpoint, it should be natural that the properties of the present model for small $\alpha$ are similar to those of the $\Lambda$CDM model.
(Possibly, the present model may be related to environment-dependent fundamental physical constants \cite{Terazawa2012}.)

\section{Generalized second law for the present model} 
\label{GSL}

In this section, we examine the generalized second law of thermodynamics for the present model.
To this end, both the power-law corrected entropy on the Hubble horizon $S_{H}$ and the entropy of matter inside the horizon $S_{m}$ are considered \cite{Sheykhi2}.
The total entropy $S_{t}$ is given as
\begin{equation}
      S_{t} = S_{H} + S_{m}  .
\label{eq:S_t0}
\end{equation}
From Eq.\ (\ref{eq:dSH}), the rate of change of $S_{H}$ is written as
\begin{equation}
\dot{S}_{H}  =  \dot{S}_{\rm{BH}}  \left [ 1-   \left ( \frac{4- \alpha}{2}  \right )  \Psi_{\alpha}  \left ( \frac{H_{0}}{H} \right )^{2-\alpha}  \right ]   .
\label{eq:dSH_2}      
\end{equation}
In contrast, the rate of change of $S_{m}$ for the present model can be calculated from the first law of thermodynamics.
From Eq.\ (\ref{dSm3}) in Appendix\ \ref{Matter}, we have
\begin{equation}
\dot{S}_{m}  =  \dot{S}_{\rm{BH}}  \left ( \frac{\alpha \Psi_{\alpha} }{2}  \right )    \left ( \frac{H_{0}}{H} \right )^{2-\alpha}    .
\label{eq:dSm_1}      
\end{equation}
For details, see Appendix\ \ref{Matter}.

It should be noted that the generalized second law for the modified Friedmann equations has been examined using the power-law corrected entropy \cite{Radicella2010,Karami2011,Sheykhi2}.
Radicella and Pav\'{o}n studied the generalized second law, based on the Clausius relation and the principle of equipartition of energy \cite{Radicella2010}.
In contrast, Karami \textit{et al.} \cite{Karami2011} and Sheykhi and Hendi \cite{Sheykhi2} used the first law of thermodynamics to discuss the generalized second law.
In this sense, the present study is similar to the latter because $\dot{S}_{m}$ is calculated from the first law of thermodynamics.
However, the cosmological equations given by Eqs.\ (\ref{FRW01_SH})--(\ref{drho_SH}) are different from the equations in those works.
Therefore, the generalized second law discussed below is slightly different from that in those works.

Using Eqs.\ (\ref{eq:S_t0})--(\ref{eq:dSm_1}), the rate of change of the total entropy is 
\begin{align}
      \dot{S}_{t}  &= \dot{S}_{H} + \dot{S}_{m}     \notag \\
                       &= \dot{S}_{\rm{BH}}  \left [ 1-   (2- \alpha)  \Psi_{\alpha}  \left ( \frac{H_{0}}{H} \right )^{2-\alpha}  \right ]   ,
\label{eq:dS_t1}
\end{align}
where $\dot{S}_{\rm{BH}} >  0$ from Eq. (\ref{eq:dSBH_2}).
To satisfy $\dot{S}_{t} >  0 $, we require
\begin{equation}
 (2- \alpha) \Psi_{\alpha}  \left ( \frac{H_{0}}{H} \right )^{2-\alpha}     <  1    .
\label{dSt_ineq_1}
\end{equation} 
Substituting Eq.\ (\ref{eq:psi_a}) into Eq.\ (\ref{dSt_ineq_1}) and using $r_{H} = c/H$ and $r_{H0} = c/H_{0}$, we have 
\begin{equation}
 \frac{\alpha (2-\alpha)}{4- \alpha}   \left ( \frac{r_H}{r_c} \right )^{2-\alpha}     <  1    .
\label{dSt_ineq00}      
\end{equation}
If $r_{c} = r_{H}$, $0 < \alpha < 4$ is obtained from Eq.\ (\ref{dSt_ineq00}).
This constraint agrees with that from an accelerating universe discussed in the previous section.
In contrast, from Eq.\ (\ref{eq:dSH_ineq00}), $0 < \alpha < 2$ is required for $\dot{S}_{H} >0$, as examined in Sec.\ \ref{Power-law}.
The strict constraint, $0 < \alpha < 2$, is used in the present paper.
Multiplying Eq.\ (\ref{dSt_ineq_1}) by a positive value $H^{2} /(2- \alpha)$ gives 
\begin{equation}
 \Psi_{\alpha}   H_{0}^{2-\alpha} H^{\alpha}   <  \frac{  H^{2} }{ 2 - \alpha}  , 
\label{dSt_ineq_2}
\end{equation} 
and using Eq.\ (\ref{eq:fa}), we obtain
\begin{equation}
f_{\alpha} (H)     <  \frac{  H^{2} }{2 - \alpha}    .
\label{dSt_ineq_3}
\end{equation} 
The inequality given by Eq.\ (\ref{dSt_ineq_3}) indicates that the extra driving term $f_{\alpha} (H)$ is restricted by the generalized second law of thermodynamics, i.e., $\dot{S}_{t} >0$.
Keep in mind that $f_{\alpha} (H)  <  \frac{ 2 H^{2} }{4 - \alpha} $ is obtained directly from $\dot{S}_{H} >0$, without using $\dot{S}_{t} >0$.
In this way, the second law of thermodynamics can constrain the value of an extra driving term because a power-law corrected entropy is employed in the present study.

Numerous observations imply $\dot{H}<0$ \cite{Krishna2017}, as discussed in Sec.\ \ref{Bekenstein-Hawking entropy}.
Therefore, when $H= H_{0}$, the right-hand side of Eq.\ (\ref{dSt_ineq_3}) is a minimum.
The strictest constraint can be written as
\begin{equation}
 f_{\alpha} (H)     <  \frac{ H_{0}^{2} }{2 - \alpha}    .
\label{dSt_Cond_1}
\end{equation}
Excluding a case of $\alpha \approx 2$ and using $O ( \frac{1}{2-\alpha} )  \approx 1$, the order of the extra driving term, $f_{\alpha} (H)$, can be approximately written as
\begin{equation}
  O  ( f_{\alpha} (H)  )  \lessapprox     O \left (  H_{0}^{2}   \right )       .
\label{dSt_Cond_2}
\end{equation}
The constraint given by Eq.\ (\ref{dSt_Cond_2}) is derived from the generalized second law of thermodynamics, $\dot{S}_{t} >0$.
As a matter of fact, Eq.\ (\ref{dSt_Cond_2}) can be obtained directly from $f_{\alpha} (H)     <  \frac{ 2 H^{2} }{4 - \alpha} $, which is based on $\dot{S}_{H} >0$.
That is, the equivalent constraint can be obtained from $\dot{S}_{H} >0$, without using $\dot{S}_{t} >0$.
(A similar constraint was discussed in Ref.\ \cite{Koma10}. However, it was derived from a mathematical condition to obtain a constant-like term, unlike in the present study.)

In the $\Lambda$CDM model, the order of the density parameter $\Omega_{\Lambda}$ for $\Lambda$ is $1$, e.g., $\Omega_{\Lambda} =0.692$ from the Planck 2015 results \cite{Planck2015}.
Accordingly, the order of the cosmological constant term, $\Lambda /3$, in the Friedmann and acceleration equations, can be approximately written as 
 \begin{equation}
  O \left ( \frac{\Lambda}{3}  \right )   = O \left ( \Omega_{\Lambda}  H_{0}^{2}   \right )  \approx   O  \left ( H_{0}^{2}  \right ) , 
\label{L_order_0}
\end{equation} 
where $\Omega_{\Lambda}$ is defined by $\Lambda /(3 H_{0}^{2})$ \cite{Koma10}.
From Eqs.\ (\ref{dSt_Cond_2}) and (\ref{L_order_0}), the order of $f_{\alpha} (H)$ is found to be consistent with the order of $\Lambda /3$ measured in cosmological observations. 
This result may imply that the cosmological constant problem could be discussed from a thermodynamics viewpoint.

So far, we have focused on the late universe. 
We now briefly consider the inflation of the early universe. 
It is well-known that higher exponents such as $H^4$ terms are required for inflation. 
The higher exponent has been closely examined in $\Lambda (t)$CDM models, see, e.g., Ref.\ \cite{Sola_2015_2}.
In this study, an extra driving term proportional to $H^{\alpha}$ is derived from the holographic equipartition law with a power-law corrected entropy.
It is found that $0 < \alpha < 4$ is obtained from an accelerating universe (and $\dot{S}_{t} >0$), whereas $0 < \alpha < 2$ results from $\dot{S}_{H} >0$.
Therefore, in the present model, the higher exponent for inflation is likely to be restricted by the latter constraint.
The constraint on $\alpha$ has been closely studied in Ref.\ \cite{Radicella2010}.
However, cosmological equations in Ref.\ \cite{Radicella2010} are different from those in the present study.
Further studies should be required.
This task is left for future research.

\section{Conclusions}
\label{Conclusions}

We have applied a power-law corrected entropy based on a quantum entanglement to Padmanabhan's holographic equipartition law to thermodynamically examine an extra driving term in the cosmological equations for a flat FRW universe at late times.
Because of a deviation from the Bekenstein-Hawking entropy, an extra driving term (proportional to $H^{\alpha}$) in the acceleration equation can be derived from the holographic equipartition law. 
Interestingly, the obtained driving term in the acceleration equation is found to be restricted by the second law of thermodynamics.
The thermodynamic constraint indicates that the order of the driving term is consistent with the order of the cosmological constant measured by observations.
When $\alpha$ is small (i.e., when the deviation from the Bekenstein-Hawking entropy is small), the extra driving term is found to be constant-like as if it is a cosmological constant.
The small value of $\alpha$ could be interpreted as a weak quantum entanglement in the late universe. 
Therefore, it may be possible to discuss the so-called cosmological constant problem from a thermodynamics viewpoint, using the holographic equipartition law with the power-law corrected entropy for the weak quantum entanglement. 
In this way, the present study is expected to provide new insights into cosmological models from a thermodynamics viewpoint.
Keep in mind that the obtained results depend on the choice of the entropy on the horizon.

Note that the generalized second law of thermodynamics, $\dot{S}_{t} = \dot{S}_{H} + \dot{S}_{m} >0$, has been used here in discussing a thermodynamic constraint.
However, an equivalent constraint can be obtained directly from $\dot{S}_{H} >0$, without using $\dot{S}_{t} >0$.


\appendix

\section{Entropy of matter}
\label{Matter}

In this appendix, we examine the rate of change of the entropy of matter inside the horizon.
For this purpose, the first law of thermodynamics for non-adiabatic processes \cite{Prigogine_1988-1989} is briefly reviewed, according to Ref.\ \cite{Koma9}.
First, consider a closed system containing a constant number of particles in a volume $V$. From the first law of thermodynamics, the heat flow $dQ$ across a region during a time interval $dt$ is given by 
\begin{equation}
   dQ = dE  + p dV    ,  
\label{eq:ClosedFirstLaw_0}
\end{equation}
where $dE$ and $dV$ are changes in the internal energy $E$ and volume $V$ of the region, respectively \cite{Ryden1}. 
Dividing this equation by $dt$ and calculating several operations \cite{Koma9}, we have  
\begin{align}
 \frac{dQ}{dt} &= \frac{d}{dt} ( \rho c^{2}    V)   + p \frac{dV}{dt}          \notag \\
                  &= \left [ \dot{\rho} + 3  \frac{\dot{a}}{a} \left ( \rho   +  \frac{p}{c^2} \right )  \right ] c^2  \left ( \frac{4 \pi}{3} r^3  \right )   .      
\label{eq:dQdt}                            
\end{align}
In this calculation, we consider a sphere of arbitrary radius  expanding along with the universal expansion \cite{Modak2012}.
For details, see Ref.\ \cite{Koma9}.
In the following, $r$ is set to be $r_{H}$.
In addition, $dQ/dt$ is assumed to be related to reversible entropy \cite{Prigogine_1998}, and the entropy change $dS$ is assumed to be expressed by  
\begin{equation}
      dS = \frac{dQ}{T}    .
\label{eq:dS_dQT}
\end{equation}
If adiabatic (and isentropic) processes are considered, the continuity equation given by $ \dot{\rho} +  3 (\dot{a}/a) ( \rho   +  p/c^2 ) = 0 $ is obtained from Eq.\ (\ref{eq:dQdt}).
In the present model, the continuity equation is given by Eq.\ (\ref{drho_SH}), assuming energy flows across the Hubble horizon, because the present model is considered to be a particular case of $\Lambda (t)$CDM models.
The energy flow is small when the extra driving term $f_{\alpha} (H)$ is constant-like.

The temperature of matter, $T_{m}$, is assumed to be equivalent to the temperature on the Hubble horizon, $T$, given by Eq.\ (\ref{eq:T0}) \cite{Sheykhi2}.
That is, energy flows across the horizon are considered to be small.
Therefore, from Eqs.\ (\ref{eq:dQdt}) and (\ref{eq:dS_dQT}), the rate of change of entropy of matter can be written as
\begin{align}
     \dot{S}_{m} 
                      &=  \frac{1}{T} \left [ \dot{\rho} + 3  \frac{\dot{a}}{a} \left ( \rho   +  \frac{p}{c^2} \right )  \right ] c^2  \left ( \frac{4 \pi}{3} r_{H}^3  \right )   .
\label{dSm1}
\end{align}
In $\Lambda (t)$CDM models, small energy flows are favored \cite{Freese-Mimoso_2015,Sola_2009-2015,Sola_2015L14}, as discussed in Sec.\ \ref{Power-law entropy and holographic equipartition law}.
Thus, the small energy flow considered here is consistent with those works.

\begin{figure} [t]  
\begin{minipage}{0.49\textwidth}
\begin{center}
\scalebox{0.32}{\includegraphics{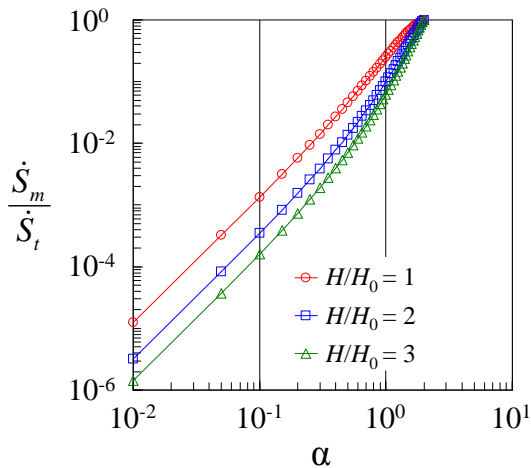}}
\end{center}
\end{minipage}
\caption{ (Color online). Dependence of $\dot{S}_{m}/\dot{S}_{t}$ on $\alpha$ 
for $H/H_{0} =1$, $2$, and $3$, plotted from $\alpha =10^{-2}$ to $\alpha =2$. Note that $r_{c}$ is set to be $r_{c} = r_{H0}$. }
\label{Fig-dSm_dSt-A}
\end{figure}

Substituting Eqs.\ (\ref{eq:rH}), (\ref{eq:T0}), and (\ref{drho_SH}) into Eq.\ (\ref{dSm1}) and using Eq.\ (\ref{eq:K-def}), we have 
\begin{align}
     \dot{S}_{m} &= \frac{1}{\frac{ \hbar H}{   2 \pi  k_{B}  }   } \left [    - \frac{3 \dot{f}_{\alpha} }{8 \pi G}   \right ] c^2  \left ( \frac{4 \pi}{3} \frac{c^3}{H^3}  \right )   
                        =    \frac{- \dot{f}_{\alpha} }{H^4} \left (  \frac{  \pi  k_{B}  c^5 }{ \hbar G }   \right )   \notag \\
                     &= \frac{- \dot{f}_{\alpha}  K }{H^4}   .
\label{dSm2}
\end{align}
In addition, substituting Eq.\ (\ref{eq:fa}) into Eq.\ (\ref{dSm2}) and using Eq.\ (\ref{eq:dSBH}), we obtain 
\begin{align}
     \dot{S}_{m} &=  \frac{- \frac{d}{dt} (\Psi_{\alpha} H_{0}^{2- \alpha}  H^{\alpha})  K }{H^4}                                                                        \notag \\
                     &=  \frac{-2K \dot{H} }{H^{3}}  \left ( \frac{ \alpha \Psi_{\alpha} }{2} \right )   \left ( \frac{ H_{0}}{H} \right )^{2- \alpha}          \notag \\
                     &=    \dot{S}_{\rm{BH}}  \left ( \frac{ \alpha \Psi_{\alpha} }{2} \right )   \left ( \frac{ H_{0}}{H} \right )^{2- \alpha}   .
\label{dSm3}
\end{align}
It can be confirmed that $\dot{S}_{m} > 0$ is satisfied because $\dot{S}_{\rm{BH}}$, $H$, $\alpha$, and $\Psi_{\alpha}$ are positive.

To observe contributions of $\dot{S}_{m}$, the dependence of $\dot{S}_{m}/\dot{S}_{t}$ on $\alpha$ is plotted in Fig.\ \ref{Fig-dSm_dSt-A}.
Here the rate of change of the total entropy, $\dot{S}_{t} = \dot{S}_{H} + \dot{S}_{m}$, is given by Eq.\ (\ref{eq:dS_t1}).
In addition, $r_{c}$ included in $\Psi_{\alpha}$ is set to be $r_{c} = r_{H0}$.
As shown in Fig.\ \ref{Fig-dSm_dSt-A}, $\dot{S}_{m}/\dot{S}_{t}$ rapidly decreases with decreasing $\alpha$.
The contribution of $\dot{S}_{m}$ is sufficiently small when $\alpha < 1$.
The small value of $\alpha$ is consistent with small energy flows. In this case, $\dot{S}_{H}$ rather than $\dot{S}_{m}$ is dominant in $\dot{S}_{t}$.

\begin{acknowledgements}
The author wishes to thank the anonymous referee for very valuable comments which improve the paper.
\end{acknowledgements}


\begin{thebibliography}{99}

%
\bibitem{PERL1998_Riess1998} S. Perlmutter \textit{et al.},  Nature (London) \textbf{391}, 51 (1998); A. G. Riess \textit{et al.}, Astron. J. \textbf{116}, 1009 (1998). 


\bibitem{Planck2015} P. A. R. Ade \textit{et al.}, Astron. Astrophys. \textbf{594}, A13 (2016).     

\bibitem{Weinberg1989} 
S. Weinberg, Rev. Mod. Phys. \textbf{61}, 1 (1989);  
I. Zlatev, L. Wang, P. J. Steinhardt, Phys. Rev. Lett.  \textbf{82}, 896 (1999); 
V. Sahni, A. A. Starobinsky, Int. J. Mod. Phys. D \textbf{9}, 373 (2000); 
S. M. Carroll, Living Rev. Relativity \textbf{4}, 1 (2001); 
T. Padmanabhan, Phys. Rep. \textbf{380}, 235 (2003);
J. D. Barrow, D. J. Shaw, Phys. Rev. Lett. \textbf{106}, 101302, (2011).



\bibitem{Prigogine_1988-1989}  
I. Prigogine, J. Geheniau, E. Gunzig, P. Nardone, Proc. Natl. Acad. Sci. U.S.A. \textbf{85}, 7428 (1988). 

\bibitem{Lima-Others1996-2016}   
J. A. S. Lima, A. S. M. Germano, L. R. W. Abramo, Phys. Rev. D \textbf{53}, 4287 (1996); 
W. Zimdahl, D. J. Schwarz, A. B. Balakin, D. Pav\'{o}n, Phys. Rev. D \textbf{64}, 063501 (2001); 
T. Harko, Phys. Rev. D \textbf{90}, 044067 (2014); 
J. A. S. Lima, R. C. Santos, J. V. Cunha, J. Cosmol. Astropart. Phys. 03 (2016) 027.



\bibitem{Freese-Mimoso_2015}
K. Freese, F. C. Adams, J. A. Frieman, E. Mottola, Nucl. Phys. \textbf{B287}, 797 (1987); 
J. M. Overduin, F. I. Cooperstock, Phys. Rev. D \textbf{58}, 043506 (1998); 
J. P. Mimoso, D. Pav\'{o}n, Phys. Rev. D \textbf{87}, 047302 (2013).


\bibitem{Sola_2009-2015} 
S. Basilakos, M. Plionis, J. Sol\`{a}, Phys. Rev. D \textbf{80}, 083511 (2009);
J. Grande, J. Sol\`{a}, S. Basilakos, M. Plionis, J. Cosmol. Astropart. Phys. 08 (2011) 007; 
A. G\'{o}mez-Valent, J. Sol\`{a}, S. Basilakos, J. Cosmol. Astropart. Phys. 01 (2015) 004.
%


\bibitem{Nojiri2006}  
S. Nojiri, S. D. Odintsov, Phys. Lett. B \textbf{639}, 144 (2006); 
Y. Wang, D. Wands, G.-B. Zhao, L. Xu, Phys. Rev. D \textbf{90}, 023502 (2014); 
N. Tamanini, Phys. Rev. D \textbf{92}, 043524 (2015); 
%
Q. Wang, Z. Zhu, W. G. Unruh, Phys. Rev. D \textbf{95}, 103504 (2017).

%




\bibitem{Bamba} 
 K. Bamba, S. Capozziello, S. Nojiri, S. D. Odintsov, Astrophys. Space Sci. \textbf{342}, 155 (2012).












\bibitem{Hooft-Bousso}
G. 't Hooft, arXiv:gr-qc/9310026; L. Susskind, J. Math. Phys. \textbf{36}, 6377 (1995); R. Bousso, Rev. Mod. Phys. \textbf{74}, 825 (2002).


















\bibitem{Sheykhi1}
A. Sheykhi, Phys. Rev. D \textbf{81}, 104011 (2010);
K. Karami, A. Sheykhi, N. Sahraei, S. Ghaffari, Eur. Phys. Lett. \textbf{93}, 29002 (2011).


\bibitem{Sadjadi1} H. M. Sadjadi, M. Jamil, Eur. Phys. Lett. \textbf{92}, 69001 (2010); 
%
S. Mitra, S. Saha, S. Chakraborty, Mod. Phys. Lett. A \textbf{30}, 1550058 (2015).








\bibitem{Jacob1995} T. Jacobson, Phys. Rev. Lett. \textbf{75}, 1260 (1995).
%
\bibitem{Padma1}  T. Padmanabhan, Mod. Phys. Lett. A \textbf{25}, 1129 (2010).
\bibitem{Verlinde1} E. Verlinde, J. High Energy Phys. 04 (2011) 029.
%








\bibitem{Easson12}  D. A. Easson, P. H. Frampton, G. F. Smoot, Phys. Lett. B \textbf{696}, 273 (2011); Int. J. Mod. Phys. A \textbf{27}, 1250066 (2012).
\bibitem{Koivisto1Basilakos1-Gohar}
Y. F. Cai, J. Liu, H. Li,      Phys. Lett. B \textbf{690}, 213 (2010); 
T. S. Koivisto, D. F. Mota, M. Zumalac\'{a}rregui, J. Cosmol. Astropart. Phys. 02 (2011) 027; 
S. Basilakos, J. Sol\`{a}, Phys. Rev. D \textbf{90}, 023008 (2014); 
M. P. D\c{a}browski, H. Gohar, V. Salzano, Entropy \textbf{18}, 60 (2016).


\bibitem{Koma4}  N. Komatsu, S. Kimura, Phys. Rev. D \textbf{87}, 043531 (2013); N. Komatsu, JPS Conf. Proc. \textbf{1}, 013112 (2014).
\bibitem{Koma5-8}  N. Komatsu, S. Kimura, Phys. Rev. D \textbf{88}, 083534 (2013); 
\textbf{89}, 123501 (2014);
\textbf{90}, 123516 (2014);
\textbf{92}, 043507 (2015).
\bibitem{Koma9}  N. Komatsu, S. Kimura, Phys. Rev. D \textbf{93}, 043530 (2016).











%
\bibitem{Bekenstein1Hawking1}  
J. D. Bekenstein, Phys. Rev. D \textbf{7}, 2333 (1973); Phys. Rev. D \textbf{9}, 3292 (1974);  Phys. Rev. D \textbf{12}, 3077 (1975); 
S. W. Hawking, Phys. Rev. Lett. \textbf{26}, 1344 (1971); Nature \textbf{248}, 30 (1974); Commun. Math. Phys. \textbf{43}, 199 (1975); Phys. Rev. D \textbf{13}, 191 (1976).
%





\bibitem{Padma2012A}  T. Padmanabhan, arXiv:1206.4916 [hep-th].
\bibitem{Padma2012}  T. Padmanabhan, Res. Astron. Astrophys. \textbf{12}, 891 (2012).
%
%
\bibitem{Cai2012-Tu2013}                                                                               
R. G. Cai, J. High Energy Phys. 1211 (2012) 016; 
K. Yang, Y.-X. Liu, Y.-Q. Wang, Phys. Rev. D \textbf{86} 104013 (2012); 
A. Sheykhi,  M. H. Dehghani, S. E. Hosseini, Phys. Lett. B \textbf{726}, 23 (2013).
%


\bibitem{Tu2013-2015}  
Fei-Quan Tu, Yi-Xin Chen, J. Cosmol. Astropart. Phys. 05 (2013) 024; 
%
E. Chang-Young, D. Lee, J. High Energ. Phys. 04 (2014) 125; 
S. Chakraborty, T. Padmanabhan, Phys. Rev. D \textbf{92}, 104011 (2015).
%



\bibitem{Padma2014-2015}  
T. Padmanabhan, H. Padmanabhan, Int. J. Mod. Phys. D \textbf{23}, 1430011 (2014); 
T. Padmanabhan, Mod. Phys. Lett. A \textbf{30}, 1540007 (2015).


\bibitem{ZLWang2015}  
Zi-Liang Wang, Wen-Yuan Ai, Hua Chen, Jian-Bo Deng, Phys. Rev. D \textbf{92}, 024051 (2015).
%
\bibitem{Tu2015}  
Fei-Quan Tu, Yi-Xin Chen, Gen. Relativ. Gravit. \textbf{47}, 87 (2015).

\bibitem{Yuan2013} Fang-Fang Yuan, Yong-Chang Huang, arXiv:1304.7949v3.


\bibitem{Krishna2017} Krishna P B, Titus K Mathew, arXiv:1702.02787v1.





\bibitem{Koma10}  N. Komatsu, Eur. Phys. J. C \textbf{77}, 229 (2017).







\bibitem{Tsa1}    C. Tsallis,  {\it Introduction to Nonextensive Statistical Mechanics: Approaching a Complex World} (Springer, New York, 2009).
\bibitem{Tsa0}    C. Tsallis, J. Stat. Phys. \textbf{52},  479 (1988).
\bibitem{Ren1}    A. R\'{e}nyi, \textit{Probability Theory} (North-Holland, Amsterdam, 1970).

\bibitem{Tsallis2012}  C. Tsallis, L. J. L. Cirto, Eur. Phys. J. C \textbf{73}, 2487 (2013). 



\bibitem{Czinner1}       T. S. Bir\'{o}, V. G. Czinner, Phys. Lett. B \textbf{726}, 861 (2013).
\bibitem{Czinner2}       V. G. Czinner, H. Iguchi, Phys. Lett. B \textbf{752}, 306 (2016).








%
\bibitem{LQG2004_1} A. Chatterjee, P. Majumdar, Phys. Rev. Lett. \textbf{92}, 141301 (2004).
\bibitem{LQG2004_2} A. Ghosh, P. Mitra, Phys. Rev. D \textbf{71}, 027502, (2005).
\bibitem{LQG2004_3} K.A. Meissner, Class. Quantum Grav. \textbf{21}, 5245, (2004).




%
\bibitem{Das2008} S. Das, S. Shankaranarayanan, S. Sur, Phys. Rev. D \textbf{77}, 064013 (2008).

\bibitem{Radicella2010} N. Radicella, D. Pav\'{o}n, Phys. Lett. B \textbf{691}, 121 (2010).

\bibitem{Karami2011} K. Karami, A. Abdolmaleki, Z. Safari, S. Ghaffari, J. High Energy Phys. 08 (2011) 150.

\bibitem{Sheykhi2} A. Sheykhi, S. H. Hendi, Phys. Rev. D \textbf{84}, 044023 (2011). 

\bibitem{Sheykhi2011} A. Sheykhi, M. Jamil, Gen. Relativ. Gravit. \textbf{43}, 2661 (2011).
%
\bibitem{Karami2013} K. Karami, S. Asadzadeh, A. Abdolmaleki, Z. Safari, Phys. Rev. D \textbf{88}, 084034 (2013).

\bibitem{Saha2016} P. Saha, U. Debnath, Eur. Phys. J. C \textbf{76}, 491 (2016).










\bibitem{Farooq2017}  O. Farooq, F. R. Madiyar, S. Crandall, B. Ratra, Astrophys. J. \textbf{835}, 26 (2017).




\bibitem{Dvali2003} G. Dvali, M. S. Turner, arXiv:astro-ph/0301510v1.












%





\bibitem{Sola_2015L14}
J. Sol\`{a}, A. G\'{o}mez-Valent, J. C. P\'{e}rez, Astrophys. J. \textbf{811}, L14 (2015).


\bibitem{Terazawa2012} H. Terazawa, arXiv:1202.1859v13.


\bibitem{Sola_2015_2}
J. Sol\`{a}, Int. J. Mod. Phys. A \textbf{31}, 1630035 (2016).





\bibitem{Ryden1}  B. Ryden, \textit{Introduction to Cosmology} (Addison-Wesley, Reading, MA, 2002). 





\bibitem{Modak2012} S. K. Modak, D. Singleton, Phys. Rev. D \textbf{86}, 123515 (2012).



\bibitem{Prigogine_1998} D. Kondepudi, I. Prigogine, \textit{Modern Thermodynamics: From Heat Engines to Dissipative Structures} (John Wiley \& Sons, New York, 1998).



\end{thebibliography}
\end{document}